\documentclass{aa}
\usepackage{amsmath,graphicx,amssymb}
\usepackage{txfonts}
\usepackage{natbib}
\bibpunct{(}{)}{;}{a}{}{,}

\newcommand\hvezda{\object{HD 64740}}
\newcommand{\zav}[1]{\left(#1\right)}

\newcommand\intvidpo{\!\!\int\limits_{\begin{array}{c}\text{\scriptsize
visible}\\[-2mm]\text{\scriptsize surface}\end{array}}\!\!}

\newlength\staretab

\makeatletter
\def\sgn{\mathop{\operator@font sgn}\nolimits}
\makeatother

\defcitealias{bohlender88}{B88}

\begin{document}

\title{Ultraviolet and visual flux and line variations
of one of the least variable Bp stars HD 64740\thanks{Partly based on
observations obtained at the European Southern Observatory
(ESO programme 089.D-0153(A)).}}

\author{J.~Krti\v{c}ka\inst{1} \and J. Jan\'\i k\inst{1} \and
        H.~Markov\'a\inst{1} \and Z.~Mikul\'a\v sek\inst{1,2} \and
        J.~Zverko\inst{1} \and M.~Prv\'ak\inst{1} \and M.~Skarka\inst{1}} 


\institute{Department of Theoretical Physics and Astrophysics,
           Masaryk University, Kotl\'a\v rsk\' a 2, CZ-611\,37 Brno, Czech Republic
            \and
            Observatory and Planetarium of J. Palisa, V\v SB -- Technical
            University, Ostrava, Czech Republic}

\date{Received}

\abstract {The light variability of hot magnetic chemically peculiar stars is
typically caused by the flux redistribution in spots with peculiar
abundance. This raises the question why some stars with surface abundance spots show
significant
rotational light variability, while others do not.}
{We study the Bp star HD 64740 to
investigate how its
remarkable inhomogeneities in the surface distribution of helium and
silicon, and the corresponding strong variability of many spectral
lines, can result in one of the faintest photometric variabilities among the
Bp stars.}
{We used model atmospheres and synthetic spectra calculated for the silicon and
helium abundances from surface abundance maps to predict the ultraviolet and
visual light and line variability of HD 64740. The predicted fluxes and line
profiles were compared with the observed ones derived with the IUE, HST, and Hipparcos
satellites and with spectra acquired using the FEROS spectrograph at the 2.2m
MPG/ESO telescope in La Silla.} {We are able to reproduce the observed visual
light curve of HD 64740 assuming an inhomogeneous distribution of iron correlated
with silicon distribution. The light variations in the ultraviolet are hardly
detectable. We detect the variability of many ultraviolet lines of carbon,
silicon, and aluminium and discuss the origin of these lines and the nature of
their variations.} {The maximum abundances of helium and silicon on the
surface of HD 64740 are not high enough to cause significant light variations.
The detected variability of many ultraviolet lines is most likely of atmospheric
origin and reflects the inhomogeneous elemental surface distribution. The
variability of the \ion{C}{iv} resonance lines of carbon is stronger and 
it probably results from the dependence of the wind mass-loss rate on
the chemical
composition and magnetic field orientation.
We have not been able to detect a clear signature of the
matter trapped in the circumstellar clouds.}


\titlerunning{Modelling ultraviolet and visual spectral energy distribution
variations of Bp star HD 64740}

\authorrunning{J.~Krti\v{c}ka et al.}
\maketitle

\section{Introduction}

Atmospheres of stars from the upper part of the main sequence provide us with a
unique laboratory for studying the fascinating effects of the radiative
force. While in the hot luminous stars the radiative force is strong enough to
launch a line-driven wind \citep{pulvina}, the atmospheres of numerous less
luminous stars are relatively quiet. Consequently, the diffusion under the influence of
the radiative and gravitational forces causes pronounced deviations from the
typical chemical composition (\citealt{vaupreh}, \citealt{mpoprad}). Stars with
such atmospheres are called chemically peculiar.


Chemically peculiar (CP) stars constitute a very diverse group of stars, and the
deviations from the solar chemical composition is the only attribute common
to all of them.
The stars may differ in the strength of their
peculiarity \citep{briketka}, elements with peculiar abundance \citep{rompreh},
horizontal and vertical homogeneity \citep{racaver}, variability \citep{mikzoo},
stability of their rotational periods \citep{mikbra2}, and the presence and
strength of the magnetic field \citep{lamaot}, to mention the most
important differences.

Helium-rich stars are among the most enigmatic objects of the group of CP
stars. In the Hertzsprung-Russell diagram they lie at the border between 
stars with strong line-driven winds and stars with weak winds, which can be
hardly detected. The hottest ones have winds with typical mass-loss rates of
about $10^{-9}\,\text{M}_\odot\,\text{year}^{-1}$. The wind can fill stellar magnetospheres in the
presence of the magnetic field, creating
magnetospheric clouds \citep{labor,shobro,smigro,towog}.

Light, and consequently also the spectral energy distribution variability, is a
typical property of many CP stars. There are several types of light variability
observed in CP stars, including pulsations \citep{balonek1,pawraf}, or the
variability resulting from the light absorption in magnetospheric clouds
\citep{towog}. However, the most typical light variability observed in CP stars
is the rotational one, which is connected with the presence of spots on the
surface of CP stars. The flux redistribution due to the bound-bound (line) and
bound-free (continuum) transitions in the inhomogeneously distributed chemical
elements on the surface of a rotating star \citep{peter,molnar,lanko,lukor}
results in the rotational variability of a CP star. The typical amplitude of
the rotational variability is on the order of hundredths of magnitude.

Precise abundance maps \citep{kowas} enable us to successfully simulate the
light variability of CP stars. Several main sources of the rotational light
variability of CP stars were identified. Bound-free transitions of silicon and
line transitions of iron seem to be the most universal ones
\citep{krivo,myhr7224} operating in CP stars with different effective
temperatures. Other elements may contribute to the light variability in specific
spectral types, including helium in hot helium-rich stars
\citep{myhd37776}, and chromium in cooler CP stars \citep{seuma,mycuvir}.

It is interesting to investgate, why some stars display strong light
variations and others do not. The helium-rich star \hvezda\ \citep[HR
3089,][]{higas} belongs to the latter ones, and
was marked by \citet{adelpidi} as one of the least
variable stars in Hipparcos photometry.
While this designation may seem outdated in the era of space-based photometry with
micromagnitude precision \citep{balonek2}, we feel that it is still
meaningful
in comparison with stars with much higher light
variability amplitudes. Similarly to other helium-rich stars, \hvezda\ has a strong
magnetic field \citep{borla} and shows emission of circumstellar origin in the
wings of H$\alpha$ \citep{peral}. The star is an X-ray source \citep{draci}
located in an interesting part of the sky at the borders of the well-known Gum
nebula \citep{eda}.

The abundance of chemical elements in the spots on \hvezda\ surface was derived
by \citet[hereafter \citetalias{bohlender88}, see also
\citealt{bohla}]{bohlender88}. Clearly, the star \hvezda\ is an ideal target for
studying the light and spectral energy distribution variability of least-variable CP stars.

\section{Observations}

When studying periodically variable observables, such as light curve of a star,
it is important to know the period and ephemeris of the variability.
We attempted to derive a new ephemeris for the studied star using the
observations available to us, i.e., the historical photometry, and optical and
ultraviolet spectroscopy.
However, we were not successful mainly because of the lack of suitable
information about the phase in the data available to us. The
photometric data are very noisy, the UV spectroscopy covers only a limited time
interval and the optical spectroscopy is incomplete.
Consequently,
we adopted the ephemeris derived by
\citet{bolabrth}
\begin{equation}
\label{spatne}
\text{JD}=(2\,444\,611.859\pm0.042)+(1.33026\pm0.00006)\,E
\end{equation}
to calculate the rotational phase. The zero-phase of this ephemeris is
centered on the minimum of the magnetic field. 
The ephemeris is fairly precise, but the precision is not high enough to
accommodate modern data as well. This is not a problem for the key data on which
this study is based (IUE and Hipparcos observations), because they were obtained in the years 1981 -- 1992, where
the uncertainty of the phase is only about 0.1. On the other hand, the
uncertainty of the phase of the data derived after 2005 exceeds 0.3,
consequently, these
data cannot be used for any reliable phase-dependent analysis.

\subsection{UV spectroscopy}

We used IUE observations of \hvezda\ extracted from the INES
database \citep[see Table \ref{iuetab}]{ines}  using the SPLAT package
\citep[see also \citealt{pitr}]{splat}. We used high-dispersion large aperture
spectra in the domains 1150--1900~\AA\ (SWP camera) and 2000--3200~\AA\ (LWP
camera). The relatively high number of SWP spectra allowed us to study the
ultraviolet (UV) light variability of the star, but only information about
the flux distribution could be obtained from the sole LWP spectrum. The SWP
spectra are available in two spectral resolutions. In most applications we used
spectra with the higher spectral resolution of 0.2\,\AA.
The uncertainty of the phase of IUE/SWP observations is $0.04-0.05$ from the
ephemeris Eq.~\eqref{spatne}, consequently, the phase is known precisely for these
data. The uncertainity of the phase of IUE/LWP  observation is somewhat larger, about 0.1.
There are also six narrow
HST/STIS spectra available. All these spectra were obtained nearly in the
same phase, therefore we selected only one of them for our study. 
The phase of HST/STIS  observations has a too large uncertainty of 0.4 (from
Eq.~\eqref{spatne}), hence it does not 
bear any information and we did not include it in Table \ref{iuetab}.
There is
also a {\em Copernicus} spectrum available for \hvezda\ \citep{kopernik},
but its wavelength region mostly coincides with the IUE one, so we
did not use it here.

\begin{table}[t]
\caption{List of IUE and HST observations of \hvezda\ used here}
\label{iuetab}
\begin{center}
\begin{tabular}{ccccc}
\hline
Satellite & Camera & Image & Julian date &  Phase\\
        &&&   2\,400\,000+\\
\hline
IUE & SWP & 14442 &  44\,796.21075      &          0.583 \\
IUE & SWP & 14459 &  44\,798.18414      &          0.067 \\
IUE & SWP & 14472 &  44\,800.17412      &          0.563 \\
IUE & SWP & 14491 &  44\,802.17227      &          0.065 \\
IUE & SWP & 14508 &  44\,804.19079      &          0.582 \\
IUE & SWP & 14659 &  44\,822.04922      &          0.007 \\
IUE & SWP & 14686 &  44\,824.24160      &          0.655 \\
IUE & SWP & 15760 &  44\,953.89902      &          0.123 \\
IUE & SWP & 15809 &  44\,958.62916      &          0.678 \\
IUE & SWP & 19101 &  45\,362.52972      &          0.304 \\
IUE & SWP & 19104 &  45\,362.63918      &          0.386 \\
IUE & SWP & 19106 &  45\,362.70596      &          0.436 \\
IUE & SWP & 19108 &  45\,362.77153      &          0.486 \\
IUE & SWP & 19118 &  45\,363.65535      &          0.150 \\
IUE & SWP & 19121 &  45\,363.75405      &          0.224 \\
IUE & SWP & 19123 &  45\,363.81854      &          0.273 \\
IUE & SWP & 19132 &  45\,364.57750      &          0.843 \\
IUE & SWP & 19135 &  45\,364.70901      &          0.942 \\
IUE & SWP & 19138 &  45\,364.80958      &          0.018 \\
IUE & SWP & 19146 &  45\,365.55573      &          0.579 \\
IUE & SWP & 19149 &  45\,365.66782      &          0.663 \\
IUE & SWP & 19151 &  45\,365.73433      &          0.713 \\
IUE & SWP & 19153 &  45\,365.81644      &          0.774 \\
IUE & LWP & 12335 &  47\,150.62037      &          0.470 \\ 
HST & STIS &      &  55\,319.46898     &    \\
\hline
\end{tabular}
\end{center}
\end{table}

\begin{figure*}[t]
\centering \resizebox{0.95\hsize}{!}{\includegraphics{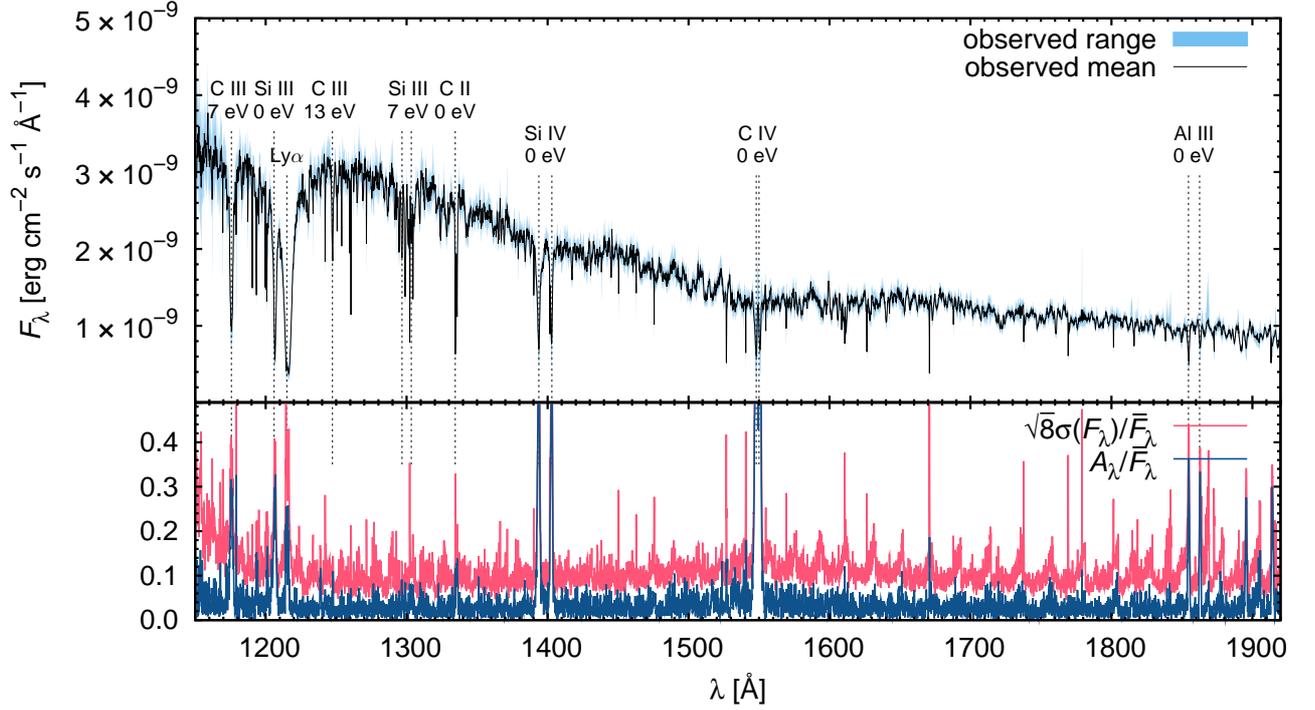}}
\caption{Observed far-ultraviolet fluxes of \hvezda\ and their variations. The
fluxes were smoothed with a Gaussian filter with a dispersion of 0.1\,\AA\ to
reduce the noise. {\em Upper panel:} Mean
flux (solid line) and the range of flux variations (lighter color).
{\em Lower
panel:} The wavelength dependence of the flux dispersion (over the rotational
period) and the effective amplitude Eq.~\eqref{alambda} divided by the mean
flux.}
\label{prtok}
\end{figure*}

In Fig.~\ref{prtok} we compare the mean observed UV flux with the minimum and
maximum one at each wavelength. Moreover, we used two different tests to detect
the line variability. We plot the dispersion of the data divided by a
mean flux at each wavelength, $\sigma(F_\lambda)/\overline{F}_\lambda$, and
the amplitude of the flux variability
\begin{equation}
\label{alambda}
A_\lambda=\sqrt{8\int_0^1\Delta F_\lambda^2(\varphi)\,\text{d}\varphi},
\end{equation}
where the difference between 
the fit by a simple polynomial $F_\lambda(\varphi)$ (with
parameters $\overline{F}_\lambda$, $a_{\lambda}$, and $b_\lambda$) and the mean
flux at a given
phase $\varphi$ is
\begin{equation}
\Delta F_\lambda(\varphi)=F_\lambda(\varphi)-\overline{F}_\lambda=
a_{\lambda}\cos(2\pi\varphi)+b_\lambda\sin(2\pi\varphi).
\end{equation}
The coefficients $a_\lambda$ and $b_\lambda$ are fitted for each wavelength of
the observed spectra. This gives $\Delta F_\lambda(\varphi)$, which is a
continuous function of $\varphi$ and can be integrated in
Eq.~\eqref{alambda}. With this fit we should be able to detect any variability
with a nonzero sinusoidal component.

A normalization of the dispersion by the mean flux  
$\sigma(F_\lambda)/\overline{F}_\lambda$ in Fig.~\ref{prtok} may lead to
artificial
peaks at the line positions even if the line is not variable. For a sinusoidal
signal in the limit of large uniformly distributed dataset the dispersion is
connected with amplitude as $A_\lambda=\sqrt{8}\sigma(F_\lambda)$. This can be
used to infer the reliability of the peaks in Fig.~\ref{prtok}, but in
general a visual inspection of the data (or advanced statistical techniques) is
necessary to test the significance of individual peaks.
From Fig.~\ref{prtok} the flux variability can be detected in the strongest
lines, which are
mostly resonance lines, whereas weaker lines show no strong variability. The
presence of resonance lines is an argument for the existence of the magnetically
confined circumstellar environment around \hvezda\ where only ground levels of
ions are strongly populated, and their variability suggests that the environment
is co-rotating \citep{shobro}. The varying lines originating from excited levels
also indicate atmospheric variability. The amplitude of the variations of
pseudo-continuum, which is mostly formed by numerous absorption lines, varies
with wavelength. This points to another process that contributes to the
pseudo-continuum variability in addition to the observing noise.
Higher variability in
the region of 1500--1600\,\AA\ indicates an inhomogeneous surface distribution of
iron (see Sect.~\ref{zelkap}).


\begin{table}[t]
\caption{List of new FEROS echelle spectra used here} \label{feros}
\begin{center}
\begin{tabular}{cccc}
\hline
Telescope & Julian date &  Phase & Exposure time\\
        &   2\,400\,000+ & & [s]\\
\hline
MPG/ESO &  56\,157.88494  &   0.526  & 180\\
MPG/ESO &  56\,157.89988  &   0.537  & 180\\
MPG/ESO &  56\,157.91512  &   0.548  & 180\\
MPG/ESO &  56\,157.93037  &   0.560  & 180\\
MPG/ESO &  56\,158.88036  &   0.274  & 180\\
MPG/ESO &  56\,158.90260  &   0.291  & 180\\
MPG/ESO &  56\,158.91762  &   0.302  & 180\\
MPG/ESO &  56\,158.93276  &   0.313  & 180\\
MPG/ESO &  56\,159.87578  &   0.022  & 240\\
MPG/ESO &  56\,159.89144  &   0.034  & 240\\
MPG/ESO &  56\,159.90722  &   0.046  & 240\\
MPG/ESO &  56\,159.92286  &   0.058  & 240\\
\hline
\end{tabular}
\end{center}
\end{table}

We were unable to detect any significant rotationally modulated broadband flux
variation (see also Fig.~\ref{hpvel}) except for the variations at 1550\,\AA.
The variations at 1550\,\AA\ are caused by variable \ion{C}{iv} lines and possibly also by
numerous iron lines, indicating an inhomogeneous flux distribution of iron (see
Sect.~\ref{zelkap}).

\subsection{Optical spectroscopy}

During run 089.D-0153(A) at the La Silla Observatory (2.2m MPG/ESO telescope)
we obtained new 12 FEROS echelle spectra with a resolution $R=48\,000$.
Initial reductions of the spectra and
their conversion into 1D images were carried out by JJ, who used IRAF
(bias-subtraction and flat-field calibration).
The wavelength calibrations were based on ThAr-Ne comparison
spectra. The wavelength coverage of these spectra is $3730-7900\,\AA$. The
journal of new spectra on which this study is based is presented in
Table~\ref{feros}. The phase of individual spectroscopic observations has
according to Eq.~\eqref{spatne} a 
too large uncertainty of 0.4, therefore we used the phase only to label
individual spectra in Table~\ref{feros}.

\subsection{Photometry}


The visual light variations are hard to detect \citep[see also
Fig.~\ref{hpvel}]{adelpidi}. The amplitude of the light variability in the
Hipparcos $H_\text{p}$ filter \citep{ESA} is only about 0.004\,mag.
The minimum of the light variations
observed around the zero phase coincides with the maximum flux at 1550\,\AA.
This indicates that the inhomogeneous surface distribution of iron could be one
of the causes of the light variability. This is also supported by the
observations made by \citet{peto} in Str\"omgren $y$, that show similar light
variability.

\begin{table}[t]
\caption{List of new $U\!B{}V$ observations used here}
\label{saaophoto}
\begin{center}
\begin{tabular}{ccccc}
\hline
Date & Epoch & No. of &  Comparison\\
     & 2\,455\,990+ & obs. &&   \\ 
\hline
9.3.2012  & 6.2872--6.4615 & 36 &   all sky \\
10.3.2012 & 7.2647--7.4461 & 46 &   all sky \\
11.3.2012 & 8.2645--8.4618 & 42 &   all sky \\
12.3.2012 & 9.2590--9.4442 & 42 &   all sky \\
\hline
\end{tabular}
\end{center}
\end{table}

Two of us (JJ and MS) obtained 166 new $U\!B{}V$ photometric observations
during four
photometric nights (9.--12.3.2012) using the 20'' telescope in South
Africa Observatory with modular photometer.
All these observations were obtained
using all-sky as well as 
differential (relative to HD~63578) photometry.
Observations were reduced with the help of the reduction program
HEC22 (rel. 14), which uses non-linear formul{\ae} to
transform from the natural to the standard $U\!B{}V$ system
\citep{homo,hec22}.
The journal of new photoelectric observations on which this study is
based is presented in Table~\ref{saaophoto}.

\section{Method of calculation}

\subsection{Stellar parameters}
\label{hvezpar}

\begin{figure}[t]
\centering
\resizebox{\hsize}{!}{\includegraphics{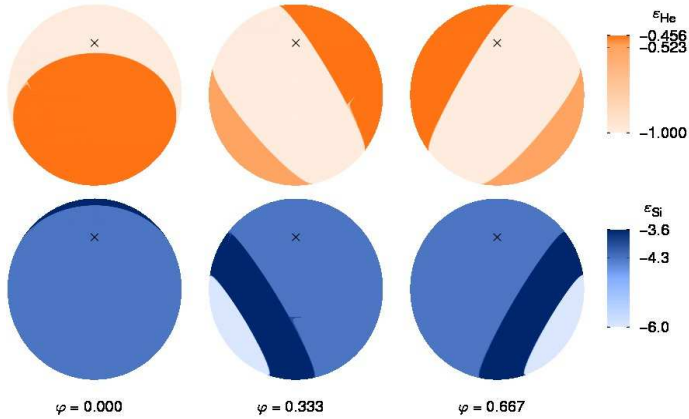}}
\caption{Distribution of helium (upper panel) and silicon (lower panel) on
the surface of \hvezda\ in three different phases.}
\label{hesipovrchm}
\end{figure}

\begin{table}[t]
\caption{Adopted \hvezda\ parameters \citepalias{bohlender88}}
\label{hvezda}
\begin{center}
\begin{tabular}{lc}
\hline
Effective temperature ${{T}_\mathrm{eff}}$ & ${24\,000\pm1000}$\,K \\
Surface gravity ${\log g}$ (cgs) & ${4.0\pm0.15}$ \\
Inclination ${i}$ & ${36^\circ\pm15^\circ}$ \\
Projection of the rotational velocity $v\sin i$ &
$140\pm10\,\text{km}\,\text{s}^{-1}$\\
Angle between rotation and magnetic axis $\beta$ & $78^\circ\pm8^\circ$ \\
Stellar mass $M$ & $11.5\pm\,2M_\odot$\\
Stellar radius $R_*$ & $6.3\pm1.8\,R_\odot$\\
Polar magnetic field strength $B_*$ & 3700\,G\\
\hline
\end{tabular}
\end{center}
\end{table}

\begin{table}[t]
\caption{Parameters of two helium spots and silicon belts.
Helium and silicon abundances were derived by
\citetalias{bohlender88}, while $\varepsilon_\text{Fe}$ is the iron
abundance assumed here.}
\label{skvrny}
\begin{center}
\begin{tabular}{lcccc}
\hline
Helium & colatitude & longitude & radius & $\varepsilon_\text{He}$\\
       & 0$^\circ$ & 0$^\circ$ & 70$^\circ$&-0.456\\
       & 160$^\circ$ & 180$^\circ$ & 60$^\circ$&-0.523\\
       & \multicolumn{3}{c}{outside the spots} & -1.0\\
\hline
Silicon & \multicolumn{2}{c}{colatitude} &
          $\varepsilon_\text{Si}$ & $\varepsilon_\text{Fe}$\\
        & \multicolumn{2}{c}{0$^\circ$--110$^\circ$}&-4.3&-4.4\\
        & \multicolumn{2}{c}{110$^\circ$--140$^\circ$}&-3.6&-3.8\\
        & \multicolumn{2}{c}{140$^\circ$--180$^\circ$}&-6.0&-5.9\\
\hline
\end{tabular}
\end{center}
\end{table}

In Table~\ref{hvezda} we list the stellar parameters and their uncertainties
estimated by \citetalias{bohlender88}, who used photometry for the effective temperature,
and fitting of hydrogen and helium lines for the surface gravity.
We used the four complete sets of $uvby\beta$ indexes listed in the SIMBAD
database and calculated the effective temperature  applying the codes
UVBYBETA \citep{modwor} and TEFFLOG \citep{smad}.
The former resulted in a $T_\text{eff}$ from 24\,300 to 25\,000\,K, while the
latter gives values from 23\,100 to 23\,700\,K with a formal error $\pm800$\,K.
The difference between the extreme values reaches nearly 2000 K, which suggests
a large uncertainity in the determination of the effective temperature, and
its influence should be taken into account.  Consciuos of this, we
considered it reasonable to adopt the set of parameters from the BSTAR2006
\citep{bstar2006} grid, which are very similar to that for which the
abundance maps were calculated.
The projected rotational velocity was derived by \citetalias{bohlender88} from helium lines.

The parameters of spots and the final adopted abundances in different regions on the \hvezda\
surface are given in Table~\ref{skvrny} (see also Fig.~\ref{hesipovrchm}).
Here the abundances are given as the logarithm of the elemental number density
$N_\text{el}$ relative to the number density of all elements $N_\text{tot}$,
$\varepsilon_\text{el}= \log\zav{N_\text{el}/N_\text{tot}}$.
The coordinates in Table~\ref{skvrny} are given with respect to the
magnetic pole.
There are two spots
with enhanced helium abundance located nearly opposite to each other with
respect to the star's center and three belts with a different silicon abundance.
There is also a nitrogen spot on the surface of \hvezda\ \citepalias{bohlender88},
but our test showed that as a result of its relatively low abundance, nitrogen
does not significantly contribute to the light curve. Consequently, we assumed
$\varepsilon_\text{N}=-4.7$, corresponding to the nitrogen abundance outside the
spots. The abundance of iron was selected to match the visual light curve, see
Sect.~\ref{predsvet}.

The stellar surface was divided in to $90\times45$ surface elements in longitude
and latitude. We calculated model atmospheres and synthetic spectra for the
chemical composition found in each surface element.

\subsection{Model atmospheres and synthetic spectra}

\begin{figure*}[t]
\centering \resizebox{0.9\hsize}{!}{\includegraphics{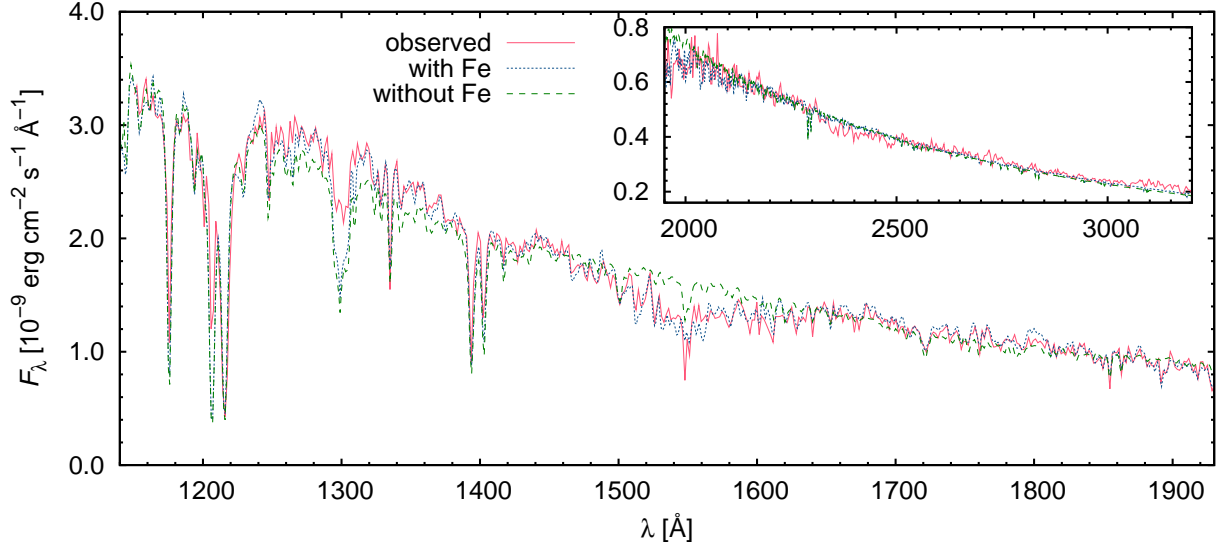}}
\caption{Comparison of observed UV flux (IUE, mean low spectral resolution
spectrum) with flux from model
atmospheres for solar iron abundance $\varepsilon_\text{Fe}=-4.6$ and for a model
without iron (the abundance of helium and silicon is $\varepsilon_\text{He}=
-1.0$, $\varepsilon_\text{Si}=-4.3$). The depression at $1550\,\AA$ is a clear
signature of iron. Both model fluxes were multiplied by a
constant factor to match the observed spectrum and convolved with a Gaussian
filter with a dispersion of 1.3\,\AA. The LWP12335 spectrum and
predicted near-UV fluxes are overplotted in the upper right.}
\label{ptokuv}
\end{figure*}

We used the code TLUSTY to calculate the NLTE plane-parallel model
atmospheres \citep{ostar2003}. The atomic data to calculate the model
atmospheres are the same as those used by \citet{bstar2006}.

To calculate the model atmospheres and synthetic spectra we assumed fixed
values of the effective temperature and surface gravity (according to
Table~\ref{hvezda}) and adopted a generic value of the microturbulent velocity
$v_\text{turb}=2\,\text{km}\,\text{s}^{-1}$. The abundance of helium, silicon,
and iron corresponds to the abundance found in individual surface regions
(see Table~\ref{skvrny}). We used the solar abundance of other elements
\citep[except nitrogen, see Sect.~\ref{hvezpar}]{asgres}. We updated the
oscillator strengths of the strong silicon lines in our line list using the data
from the Atomic Line List\footnote{http://www.pa.uky.edu/\~{}peter/atomic/}.

To predict the spectral energy distribution variations, we computed specific
intensities $I(\lambda,\theta,\varepsilon_\text{He},\varepsilon_\text{Si})$ for
$20$ equidistantly spaced values of $\mu=\cos\theta$, where $\theta$ is the
angle between the normal to the surface and the line of sight. The intensities
were calculated by the SYNSPEC code
assuming NLTE
from the TLUSTY model atmospheres.
We also took into account the same transitions as for the model
atmosphere calculations \citep[see][]{mycuvir}. Additionally, we included the
lines of all elements with the atomic number $Z\leq30$ that were not accounted
in the model atmosphere calculation.

\subsection{Phase-dependent flux distribution}
\label{vypocet}

The radiative flux in a color $c$ at the distance $D$ from the star with
radius $R_*$ is \citep{mihalas}
\begin{equation}
\label{vyptok}
f_c=\zav{\frac{R_*}{D}}^2\intvidpo I_c(\theta,\Omega)\cos\theta\,\text{d}\Omega.
\end{equation}
The integral Eq.~\eqref{vyptok} is calculated numerically over all visible
surface elements taking into account the star's rotation.
The intensity $I_c(\theta,\Omega)$ at each surface point with
spherical coordinates $\Omega$ is an integral over the specific intensity for
the appropriate abundance of helium, silicon, and iron,
\begin{equation}
\label{barint}
I_c(\theta,\Omega)=
\int_0^{\infty}\Phi_c(\lambda) \,
I(\lambda,\theta,\varepsilon_\text{He},\varepsilon_\text{Si})\, \text{d}\lambda.
\end{equation}
The response function $\Phi_c(\lambda)$ of a given filter $c$ is approximated by
a polynomial fit of tabulated values.

The magnitude difference is defined as
\begin{equation}
\label{velik}
\Delta m_{c}=-2.5\,\log\,\zav{\frac{{f_c}}{f_c^\mathrm{ref}}},
\end{equation}
where $f_c$ is calculated from Eq.~\ref{vyptok} for a given rotational phase and
${f_c^\mathrm{ref}}$ is the reference flux obtained under the
condition that the mean magnitude difference over the rotational period is zero.

\section{Mean UV fluxes and the 1550\,\AA\ iron depression}
\label{zelkap}

Iron lines substantially determine the stars' ultraviolet flux distribution 
\citep[c.f.,][]{zeminy}. This is documented in Fig.~\ref{ptokuv}, where we plot
the UV fluxes computed for the solar iron abundance, for the iron-free
atmosphere, and the observation. The predicted fluxes were scaled by
a constant factor to derive the best match between theory and observation. Both
models with and without iron agree with the observed spectrum for longer UV
wavelengths $\lambda\gtrsim 1600\,\AA$. However, in the region of
$1500-1600\,\AA$ the flux calculated without iron is significantly higher than
the observed one, while it is somewhat lower in the region of $1250-1400\,\AA$.
The flux with solar iron abundance agrees well with observations in the far-UV
region (1150--1900\,\AA), both in general shape and in many details (except for
some stronger lines, e.g., \ion{Si}{iii} lines at 1206\,\AA\ and 1300\,\AA). The
connection between the $1550\,\AA$ feature and iron was previously shown by
\citet{brows}. The flux variability at these wavelengths is a proxy for the
inhomogeneous surface distribution of iron \citep[c.f.,][]{sokold}.

Comparison of predicted near UV fluxes (2000--3200\,\AA) and one available near
UV observation in Fig.~\ref{ptokuv} shows that the flux distribution is well
reproduced, however most of minute features in the observed spectrum are not
present in the predicted spectrum. It is not clear what is the cause of this
difference. Either some line opacity is still missing in this region
due to the peculiar elemental abundances that we assume to be the solar
ones,
or the observed flux is less certain here.

\section{Predicted light variations}
\label{predsvet}

Predicted light curves were calculated from the surface abundance maps derived by
\citetalias{bohlender88} (see also Table~\ref{skvrny}) and from the emergent
fluxes computed with the SYNSPEC code, applying Eq.~\eqref{velik} for individual
rotational phases.

\begin{figure}[t]
\centering
\resizebox{0.9\hsize}{!}{\includegraphics{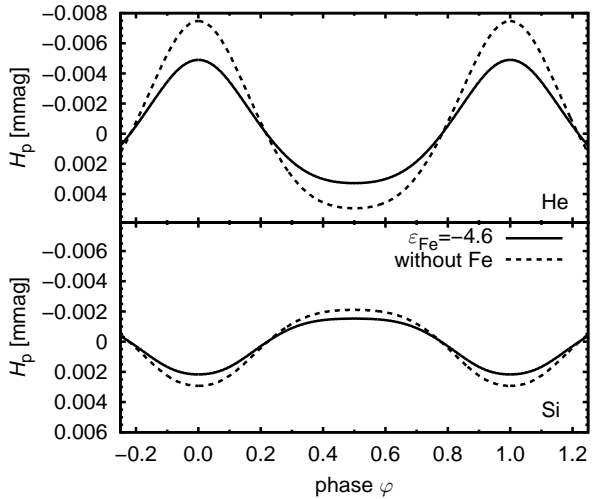}}
\caption{Predicted light variations of \hvezda\ in $H\text{p}$ photometric
filter calculated using abundance maps of one element only. The abundance of
other elements was fixed. {\em Upper plot:} Light variations due to helium
only. Silicon abundance was fixed to $\varepsilon_\text{Si}=-4.3$. {\em Lower
plot:} Light variations due to silicon only. The helium abundance was fixed to
$\varepsilon_\text{He}=-1.0$. Light curves were calculated using the solar iron
abundance ($\varepsilon_\text{Fe}=-4.6$, solid lines) and without iron (dashed
lines).}
\label{prv_hvvel}
\end{figure}

Strong far-UV opacity due to helium and silicon bound-free transitions causes
the redistribution of the flux from the far-UV region to the near-UV and visible
ones. Consequently, surface spots with enhanced helium and silicon abundance are
bright in the visible and dark in the far-UV regions \citep{myhd37776}. 
The visual light maximum is therefore observed at the phase of
the strongest lines of either helium or silicon. This is 
demonstrated in Fig.~\ref{prv_hvvel}, where we plot the light variations calculated assuming   
the inhomogeneous surface distribution of one element only, while
the other element is distributed uniformly over the whole
surface. The light variations due to helium and silicon are in antiphase,
because helium-overabundant regions are seen on the stellar surface mostly
around the phase $\varphi=0.0$, whereas regions with enhanced silicon abundance
are seen mostly around the phase $\varphi=0.5$.
Note also that one of the helium spots never dominates in the region that
directly faces the observer (see Fig.~\ref{hesipovrchm}), hence there is only one wave in the helium
light curve.

From Fig.~\ref{prv_hvvel} it follows that iron influences the
light curve even if it is homogeneously distributed on the stellar surface.
The reason for this is that iron reduces the relative opacity variations and
consequently also the flux variations caused by
other elements, therefore
the iron-free atmosphere displays a larger 
variability amplitude  than the atmosphere with a normal, solar iron abundance
distributed uniformly over the surface. 


\begin{figure}[t]
\centering
\resizebox{0.90\hsize}{!}{\includegraphics{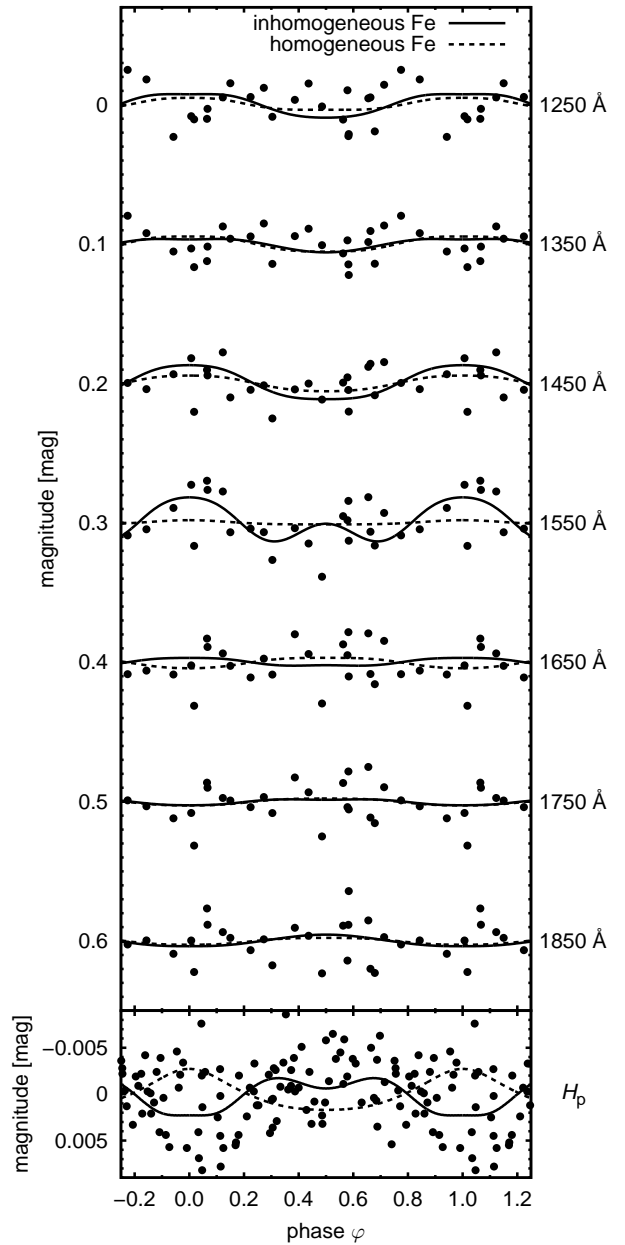}}
\caption{Predicted light variations of \hvezda\ computed taking
into account the helium and silicon surface abundance distributions after
Table~\ref{skvrny}  (solid lines: inhomogeneous iron distribution after
Table~\ref{skvrny}, dashed lines: homogeneous iron distribution) compared with IUE and HIPPARCOS
observations (dots).
{\em Upper panel:} The far ultraviolet region. Observed and predicted
fluxes were smoothed by a Gaussian filter as described in the text.
Curves for individual wavelengths were vertically
shifted. The region around \ion{C}{iv} $\lambda1548$ and $\lambda1551$ lines was
excluded from the analysis.  {\em Bottom panel:} The optical region in the
HIPPARCOS $H_\text{p}$ filter.}
\label{hpvel}
\end{figure}

Taking into account the surface distribution of helium and silicon, the
predicted light curves disagree with observed ones. The observed light maximum
occurs around the phase of the maximum strength of silicon lines, $\varphi=0.5$,
but the predicted light curve is dominated by helium, peaking around the phase
$\varphi=0.0$ (Fig.~\ref{hpvel}).
This can be deduced from Fig.~\ref{prv_hvvel}, where the amplitude of the
light variations due to helium is significantly larger than that due to
silicon.
From this we conclude that there is another element that
contributes to the light curve. From the ultraviolet observations available we
assume that this missing element is iron.

We accounted for the assumed inhomogeneous iron surface distribution in the
calculation of the light curves. Because the corresponding iron surface map is
missing, we assumed that iron has the same spot structure as silicon.
A rough
correlation between the iron and silicon abundance is typical in many CP stars
\citep[e.g.,][]{leh2}.
We tried
several different combinations of iron abundance to match the observed light
curve. The best agreement was obtained assuming that the iron abundance
scales as
$\varepsilon_\text{Fe}=\varepsilon_{\text{Fe},\odot}
\zav{\varepsilon_\text{Si}/\varepsilon_{\text{Si},\odot}}^{0.85}$, where $\odot$
denotes solar abundance of a given element. The iron abundances
calculated using this formula are given in Table~\ref{skvrny}. The
predicted $H_\text{p}$ light curve corresponds to the observed one and the
predicted minute UV flux variations do not contradict the observed flux
variations (Fig.~\ref{hpvel}). Here we applied a Gaussian function with
$\sigma=25\,$\AA\ as the
response function of the filter, centered at the corresponding
wavelengths to obtain UV broadband flux variations after Eq.~\eqref{vyptok}.
The importance of iron is especially visible
in the region of $1550\,\AA$ and in the optical region.

\begin{table}[t]
\caption{Effective amplitude Eq.~\eqref{aeff} of the light variations in
individual optical bands derived from observed and predicted light curves.}
\label{nevidim}
\begin{center}
\begin{tabular}{lccccc}
\hline
Band & $U$ & $B$ & $H_\text{p}$ & $V$ \\
Observed  $A_\text{eff}$ [mmag] & $2.1\pm1.4$ & $5.2\pm1.3$ & $3.7\pm1.1$ &
  $4.1\pm1.3$\\
Predicted $A_\text{eff}$ [mmag] & 2.9 & 4.8 & 3.7& 2.9 \\
\hline
\end{tabular}
\end{center}
\end{table}

The predicted minute light variations in other visual bands do not disagree
with observations either. This can be seen from Table~\ref{nevidim}, where we compare
the effective amplitude of the light variations $A_\text{eff}$ derived using
observed and predicted data in individual bands.
We compare only the effective amplitude and not the light curves, because the
observations are relatively noisy and the uncertainty of the phase of the
$UBV$ observations is large.
The effective amplitude is
defined here as 
\begin{equation}
\label{aeff}
A_\text{eff}=\sqrt{8\int_0^1\Delta m^2(\varphi)\,\text{d}\varphi},
\end{equation}
where $\Delta m(\varphi)$ the difference between the magnitude at a given phase
$\varphi$ and the mean amplitude. For each band $\Delta m(\varphi)$ is
derived from a simple polynomial fit
\begin{equation}
\Delta m(\varphi)=a\cos(2\pi\varphi)+b\sin(2\pi\varphi),
\end{equation}
where $a$ and $b$ are the parameters of the fit.
We note that in the Balmer continuum (in the region of Str\"omgren $u$) even
these minute changes nearly cancel, resulting in smaller amplitude in $U$ than in
$B$.

\section{Variability of UV lines}

Fig.~\ref{prtok} shows that many UV lines might be variable. The presence and
variability of \ion{Si}{iv} and \ion{C}{iv} lines in \hvezda\ spectra were
used as proof of the
circumstellar matter in the extended magnetosphere \citep{shobro}. However, hot
Bp stars show an atmospheric component of \ion{Si}{iv} and \ion{C}{iv} lines
\citep{kampa}, which may be variable if the elements are inhomogeneously
horizontally distributed. Therefore, the circumstellar matter in the
magnetosphere can be unambiguously detected only with detailed model
atmospheres.

\begin{figure}[t]
\centering
\resizebox{0.9\hsize}{!}{\includegraphics{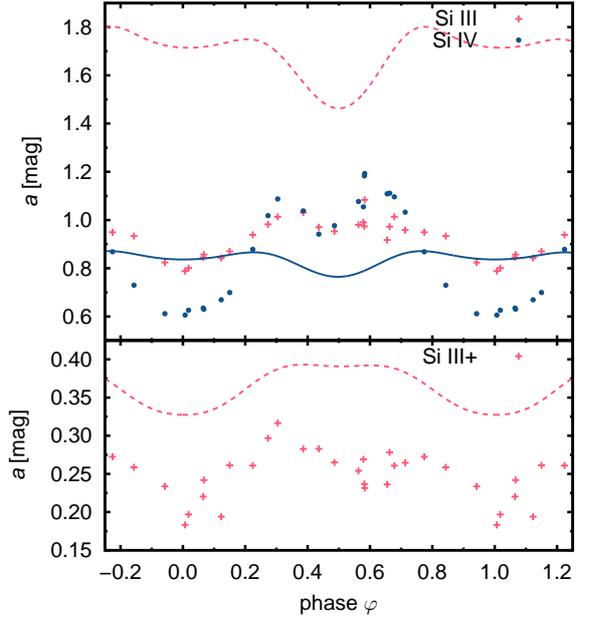}}
\caption{Observed variability of silicon line  indices Eq.~\eqref{sicarin}
(\ion{Si}{iii} crosses, \ion{Si}{iv} dots) compared with predicted ones
(\ion{Si}{iii} dashed lines, \ion{Si}{iv} solid lines).}
\label{sicar}
\end{figure}

\begin{figure*}[t]
\centering
\resizebox{0.9\hsize}{!}{\includegraphics{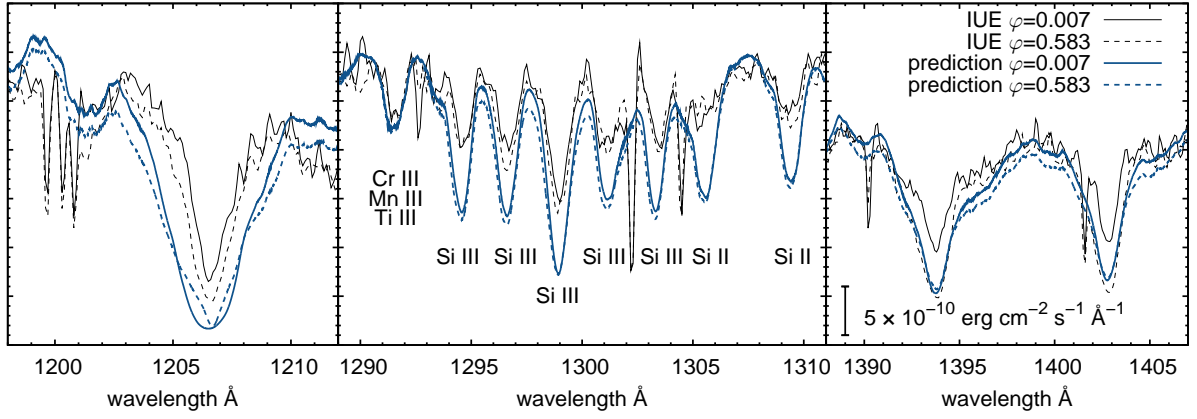}}
\caption{Observed profiles of silicon lines (black) compared with predicted
ones (blue) for phases $\varphi=0.583$ (solid lines) and $\varphi=0.007$ (dashed
lines). {\em Left:} Resonance line of \ion{Si}{iii}. {\em Middle:} Excited lines
of \ion{Si}{ii} and \ion{Si}{iii}. {\em Right:} Resonance lines of \ion{Si}{iv}.}
\label{sicarprof}
\end{figure*}

For silicon we have an abundance map at our disposal. Because the exact continuum
placement is uncertain in the UV domain, \citet{shobros} devised certain
indices to describe the line variability. Here
we used similar indices defined using the flux at the line center (expressed in
magnitudes $m_\lambda$) minus the average continuum flux,
\begin{subequations}
\label{sicarin}
\begin{align}
a(\text{\ion{Si}{iv}})&=m_{1393.7}-\frac{1}{2}\zav{m_{1388.6}+m_{1398.7}},\\
a(\text{\ion{Si}{iii}})&=m_{1205.9}-\frac{1}{2}\zav{m_{1195.8}+m_{1226.0}},\\
a(\text{\ion{Si}{iii+}})&=m_{1296.4}-\frac{1}{2}\zav{m_{1289.7}+m_{1308.2}}.
\end{align}
\end{subequations}
We calculated the line indices from low-dispersion spectra. The wavelengths in
Eq.~\eqref{sicarin} were selected from the spectra, and consequently do not
exactly match the line centers.
The comparison of the predicted and observed silicon line indices is given in
Fig.~\ref{sicar}. We detected the rotational line variability not only in
\ion{Si}{iv} lines as did \citet{shobro}, but also in the resonance and excited
lines of \ion{Si}{iii} \citep[similarly to other Bp stars,][]{shorad}. The
variability in the resonance line of \ion{Si}{iii} at 1206\,\AA\ has nearly the
same amplitude as that of \ion{Si}{iv} at 1394\,\AA\ and 1403\,\AA. 
Observed and predicted line profiles disagree
(Fig.~\ref{sicarprof}). The predicted line profiles of \ion{Si}{iii} are
significantly stronger than the observed ones, and the observed variability in the
resonance lines has a larger amplitude than the predicted one. The
observed \ion{Si}{iv} line profiles (Fig.~\ref{sicarprof}) are nicely reproduced
only in phases $0.25-0.75$.

\begin{figure}[t]
\centering
\resizebox{0.9\hsize}{!}{\includegraphics{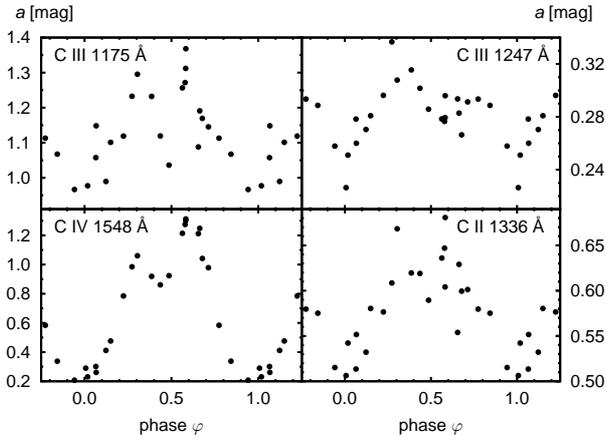}}
\caption{Observed variability of carbon line indices according to Eq.~\eqref{ccarin}.}
\label{ccar}
\end{figure}

There are many strong carbon lines in the IUE spectra. Because we do not have
any carbon abundance map at hand, we restrict ourselves to discussing 
their observed behaviour. We studied their variability using indices similar
to those in Eq.~\eqref{sicarin}
\begin{subequations}
\label{ccarin}
\begin{align}
a(\text{\ion{C}{ii} } 1336\,\AA)&=m_{1335.0}-\frac{1}{2}\zav{m_{1331.6}+m_{1340.0}},\\
a(\text{\ion{C}{iii} } 1175\,\AA)&=m_{1175.7}-\frac{1}{2}\zav{m_{1169.0}+m_{1180.8}},\\
a(\text{\ion{C}{iii} } 1247\,\AA)&=m_{1247.8}-\frac{1}{2}\zav{m_{1244.5}+m_{1249.5}},\\
a(\text{\ion{C}{iv} } 1548\,\AA)&=m_{1547.9}-\frac{1}{2}\zav{m_{1529.4}+m_{1557.9}}.
\end{align}
\end{subequations}
Many lines of different ionization states of carbon
show strong line variability, including lines originating from excited levels
(Fig.~\ref{ccar}). The maximum of the carbon line strength occurs between phases
0.3 -- 0.7. On the other hand, we were unable to detect the rotational line
variability of \ion{C}{ii} lines at $1721$~\AA\ and $1761$~\AA, probably
because of the weakness of the lines.

\begin{figure}[t]
\centering
\resizebox{0.8\hsize}{!}{\includegraphics{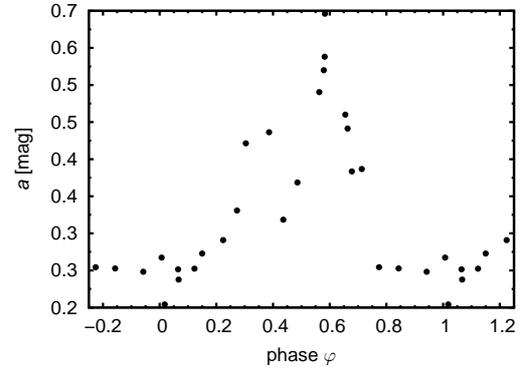}}
\caption{Observed variability of the aluminium line index according to Eq.~\eqref{alcarin}.}
\label{humpolec}
\end{figure}

\begin{figure*}[t]
\centering
\resizebox{\hsize}{!}{\includegraphics{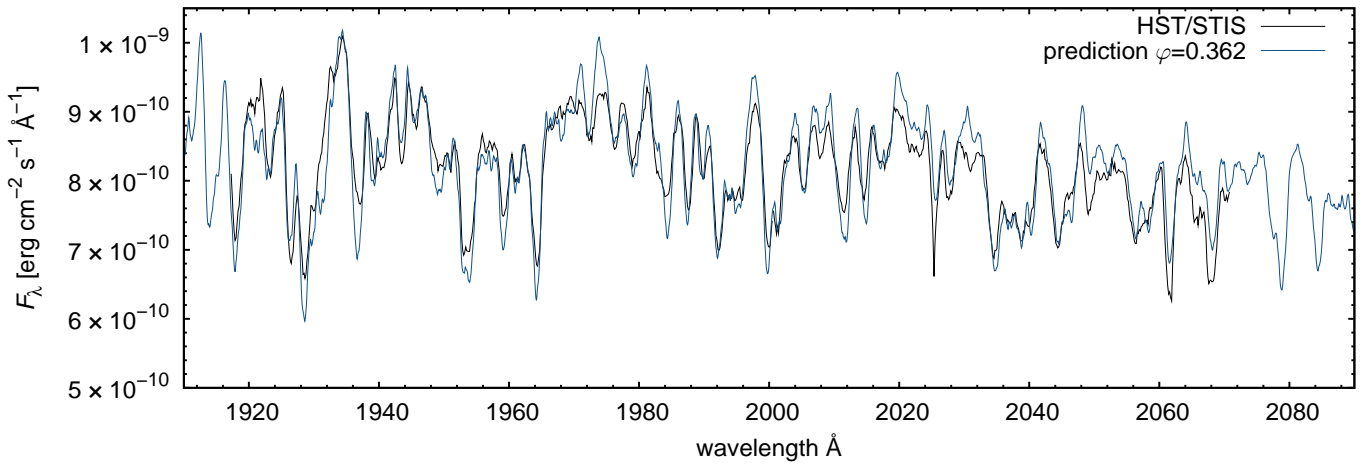}}
\caption{Comparison of the predicted and observed narrow-range spectrum.}
\label{hd64740_hst}
\end{figure*}

We also have a nitrogen abundance map available. However, the nitrogen lines
are weak in the IUE range. Consequently, we were unable to detect any
rotational line variability of the \ion{N}{ii} and \ion{N}{iii} lines. On the
other hand, resonance \ion{Al}{iii} lines show distinct line variability with
rotational period \citep[as shown for other Bp stars by][]{shorad}, as can
be seen from the plot in Fig.~\ref{humpolec} of the $1855$~\AA\ line index
defined as
\begin{equation}
\label{alcarin}
a(\text{\ion{Al}{iii}})=m_{1854.7}-\frac{1}{2}\zav{m_{1851.3}+m_{1858.0}}.
\end{equation}
Moreover, \ion{Al}{iii} resonance lines show the maximum strength around the phases
0.3 -- 0.7. 
Consequently, as can be seen seen in Figs.~\ref{sicar}, \ref{ccar}, and
\ref{humpolec}, the variability of all
the studied
lines, expressed by means of indices (Eqs.~\eqref{sicarin}--\eqref{alcarin}), show a similar
behavior, regardless of whether the line is resonant or not.

The observed and predicted HST/STIS fluxes 
in the region 1920--2070\,\AA\ agree well (see Fig.~\ref{hd64740_hst}).
All features in the spectrum are in fact blends of numerous lines, mostly of
iron.


\section{Discussion}

\subsection{Photometric variability}

Three conditions have to be met so that appreciable photometric
variability due to inhomogeneous elemental surface distribution can be
observed: i) an overabundance, ii) the amplitude of abundance variations
of the photometrically active element have to
be large enough to significantly influence the emergent flux, and
iii) the
surface spots should occupy a significant part of the stellar surface.
The maximum overabundances of helium or silicon on the \hvezda\
surface are not large enough to cause substantial flux redistribution.
Consequently, although \hvezda\ shows significant spectroscopic variations,
its photometric variability in the UV and visual bands is hardly detectable. The light
variations in stars where one of these conditions is not met have a very low
amplitude and can be detected using spacebased instruments only \citep{balonek2}.

\subsection{Are co-rotating circumstellar clouds needed?}
\label{otazka}

Bp stars are well-known for their magnetically confined circumstellar
clouds \citep{labor,nakaji,smigro,towo}. The matter lifted by the wind
from the atmosphere flows along the magnetic field lines and accumulates in the
local effective potential minima at individual field lines. The evidence for
these corotating magnetically confined circumstellar clouds is particularly
strong in stars with H$\alpha$ emission \citep[e.g.,][]{towog}. The variability
of UV lines is also frequently used as an indication of corotating clouds.
However, we argue that this evidence may be problematic as a result of
contamination of UV line profiles by atmospheric lines. 

\begin{figure*}[t]
\centering
\resizebox{0.7\hsize}{!}{\includegraphics{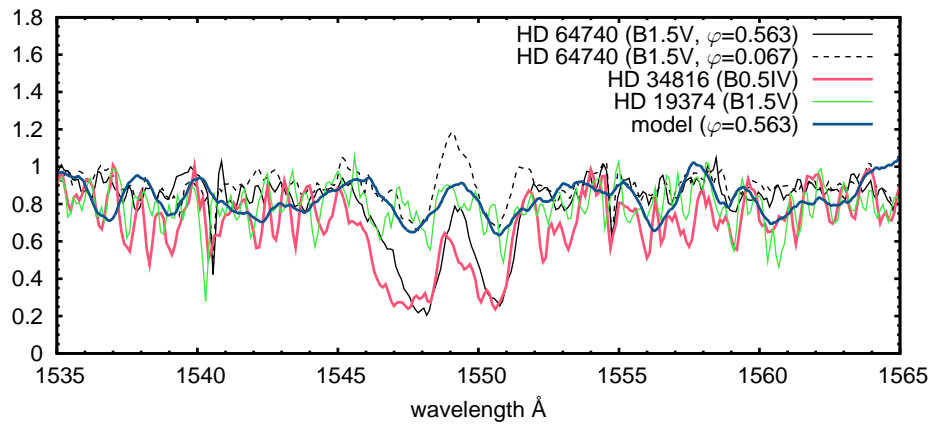}}
\caption{Comparison of \ion{C}{iv} resonance line profiles. The observed line
profiles of \hvezda\ are plotted in the phases of maximum and minimum line
strengths. The predicted line profile was calculated assuming a homogeneous carbon
abundance distribution, consequently it shows basically no variations. These
line profiles are compared with IUE line profiles of normal stars with slightly
earlier and the same spectral types than that of \hvezda. We note that the
fluxes were normalized to a pseudo-continuum given by numerous absorption lines.
}
\label{civcar}
\end{figure*}

Detailed elemental abundance maps can be used to predict the UV line profiles of
atmospheric origin and to test the nature of UV line variability. For
\hvezda\ the abundance maps are not sufficiently detailed, consequently, we had
to use indirect tests to infer the origin of the UV line variability. The matter in
the co-rotating magnetosphere accumulates in the warped disk with highest density
above the intersections of magnetic and rotational equators \citep{towo}. In
\hvezda\ this occurs for phases 0.25 and 0.75. If the additional absorption
occurs in the matter trapped in the magnetosphere, the strength of the \ion{C}{iv}
and \ion{Si}{iv} lines should be strongest at these phases and there should be
two roughly similar line strength minima at phases 0 and 0.5. However,
this is not the case; the line strengths increase at phase 0.25 and decrease at
0.75 (see Figs.~\ref{sicar}, \ref{ccar}, and \ref{humpolec}). Moreover, the
absorption lines should show profiles whose line center position slightly varies
from blue to red, while the observed line profiles only show extended blue
wings for the \ion{C}{iv} lines (Fig.~\ref{civcar}). Consequently, we
conclude that it is questionable that the additional absorption occurs in the
matter trapped in the rotating magnetosphere.

We note that although we found no clear signature of co-rotating
magnetospheric clouds, \hvezda\ most likely has such clouds.
This is documented observationally from H$\alpha$ line
profiles \citep{peral}, and also theoretically,
because the Kepler radius (at which the centrifugal force balances gravity)
$R_\text{K}=\zav{\frac{GM}{\Omega^2}}^{1/3}=1.8\,R_*$ is lower that the Alfv\'en
radius, at which the wind speed is equal to the Alfv\'en velocity,
$R_\text{A}/R_*=\zav{\frac{B_*^2R_*^2}{\dot M v_\infty}}^{1/4}=28\,R_*$
\citep{owodul}. Here we assumed a wind mass-loss rate $\dot
M=4\,\times10^{-10}\,M_\odot\,\text{year}^{-1}$ and a wind terminal velocity
$v_\infty=1700\,\text{km}\,\text{s}^{-1}$, as predicted from NLTE wind models
\citep{cmf}. The typical spindown time as a result of angular momentum loss via
the magnetized stellar wind is about 6\,Myr \citep{udowot}, which can be
measurable with a sufficiently long time span \citep{mikbra2}.

We also tested if the accumulation of numerous iron lines might
mimic absorption lines of \ion{C}{iv}, as suggested by \citet{huhas}. For
normal stars we confirm that the presence of absorption due to atmospheric
\ion{C}{iv} lines and numerous iron lines does not necessarily imply the
existence of strong outflow, as shown in Fig~\ref{civcar}. However, these lines
cannot explain the strength of absorption observed in \hvezda.

\begin{figure*}[t]
\centering
\resizebox{\hsize}{!}{\includegraphics{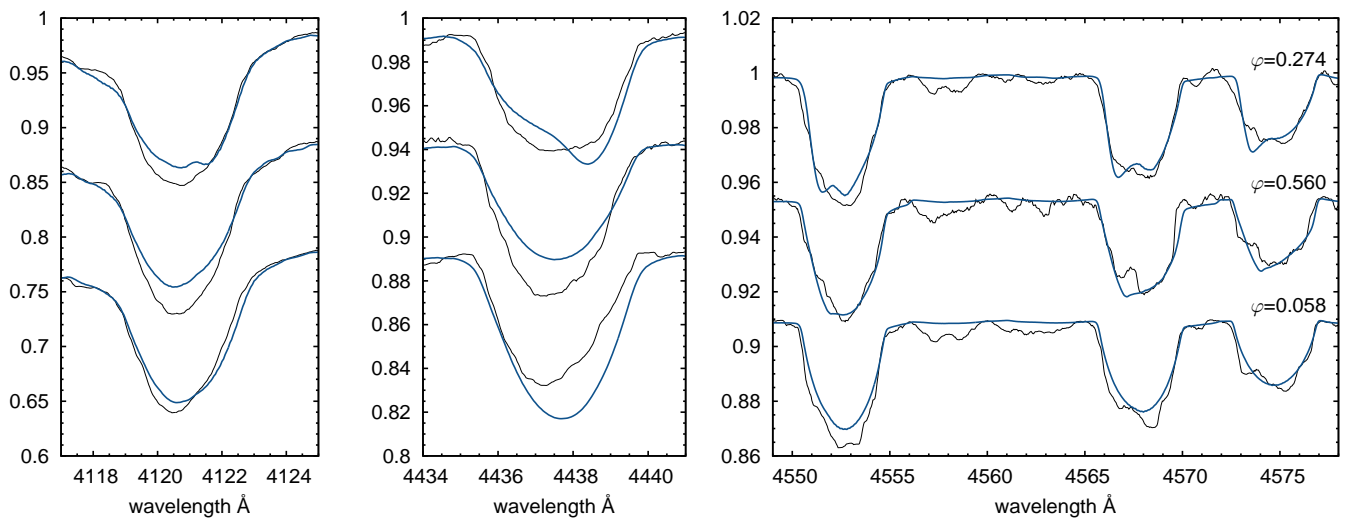}}
\caption{Line profiles of individual helium and silicon lines in different
phases (observed with FEROS) compared with line profiles
predicted using abundance maps.
Line
profiles were vertically shifted to better demonstrate the line variability.
{\em Left and middle:} Helium lines. {\em Right: Silicon lines}.}
\label{vidcar}
\end{figure*}

The strength and profile of \ion{C}{iv} lines in \hvezda\ does not correspond to
its spectral type B1.5V, but to an earlier spectral type B0.5V (see
Fig.~\ref{civcar} and \citealt{masob}). The extended blueshifted absorption in
the spectra of early spectral types is the absorption part of the P Cygni profile
originating in the stellar wind \citep[e.g.,][]{caslam}. Consequently, the
strong \ion{C}{iv} lines may originate in the stellar wind.
In Bp stars 
several effects may strengthen the wind absorption lines.
First, the wind mass-loss rate increases with
metallicity \citep[in O stars as $\dot M\sim Z^{0.67}$,][]{nlteii}. Moreover,
for a given mass-loss rate the line strength increases with increasing abundance
of a given element.
Mass-flux can be also affected by the geometry of the magnetic field
\citep{owoudan}.
These effects combined may explain the strong \ion{C}{iv}
lines observed in the spectra of Bp stars, and with the inhomogeneous elemental
surface distribution, also their variability.

From the variability of carbon line profiles in our FEROS optical spectra
we infer that carbon is distributed inhomogeneously on the surface of \hvezda.
\hvezda\ probably shows no enhanced carbon abundance \citep{kopernik,zborno}.
However, the strength of \ion{C}{iv} resonance lines may reflect the enhanced
mass-loss rate resulting from a higher abundance of other elements that drive the wind.
This corresponds to the fact that regions with enhanced silicon abundance are
visible in the phases 0.25--0.75, where the carbon resonance lines are also the
strongest. A similar effect may also affect the \ion{Si}{iv} lines.

\subsection{Weak UV lines of \ion{Si}{ii} and \ion{Si}{iii} }

The predicted UV \ion{Si}{ii} and \ion{Si}{iii} line profiles are significantly
stronger than the observed ones (Fig.~\ref{sicarprof}). In Sect.~\ref{otazka} we
argued that these lines and their variations probably do not originate in the
circumstellar medium. There are other arguments in favor of an atmospheric origin
of these lines. If these weak line profiles originate in the circumstellar
medium, the photospheric absorption profiles have to be filled
by the emission of circumstellar origin.
However, the typical line profiles in such a case
are different, showing emission in the wings in certain phases \citep{towog}.
Moreover, resonance and excited lines are affected by the same relative
amount, which is unlikely if the source of emission is the circumstellar medium.

The UV \ion{Si}{iii} line profiles can be better reproduced by models
with an effective temperature higher by roughly $2000\,\text{K}$. The
published
temperature determinations from the optical region seem to indicate the need for
an opposite shift of the effective temperature \citep{zboril,cidal,netopil}. 
However, because there is a relatively large uncertainty in the $T_\text{eff}$
determination (see
Sect.~\ref{hvezpar} for a detailed discussion), the higher effective temperature
can still be the reason for the disagreement
between observed and predicted UV \ion{Si}{iii} line profiles.

Alternatively, a significant reduction of silicon abundance is required to bring
the observations and theory in an agreement. The legitimacy of such an
abundance reduction can  in principle be tested using our optical FEROS spectra.
However, as a result of the uncertainty of the adopted ephemeris we are unable
to derive the reliable information about the phase for the FEROS observations.
Consequently, if the abundances adopted by us are correct, the predicted and
observed line profiles may differ in shape, but the line widths should be
similar. This can be seen in Fig.~\ref{vidcar}. The
abundance reduction therefore cannot explain the weak UV lines of silicon.
We also note that the line profiles calculated by \citetalias{bohlender88}
from the maps agree with their observations.

The fact that the observed UV silicon lines that originate in
the upper atmosphere are weaker than the predicted ones (while the remaining weak
optical lines are
not) may indicate a vertical silicon abundance gradient in the atmosphere
\citep[c.f.,][]{shagla,baila}.
Our numerical test have shown that a significant reduction of the silicon abundance
for the Rosseland optical depths $\tau_\text{Ross}\leq0.01$ improves the
agreement with UV line profiles of \ion{Si}{ii} and \ion{Si}{iii} while keeping
the optical silicon line profiles nearly intact. The
\ion{Si}{iv} doublet at 1394\,\AA\ and 1403\,\AA\ is also to a lesser extent affected
by the abundance stratification, but it is not clear if there is such a
horizontal and vertical silicon distribution in the atmosphere that would fit
all observations.

The adopted rather simple abundance maps are another possible source of the
differences between theory and observations. The surface abundance maps are
typically much more complex than those used here, and more detailed
maps might possibly improve the agreement between theoretical and
observed silicon line profiles.

\section{Conclusions}

We studied the UV and optical light variability of a Bp star with a very low
amplitude of the light variability, \hvezda. We have shown that the maximum
abundance of helium and silicon on the surface of \hvezda\ is not high enough to
cause a substantial light variability.

The observed and predicted UV flux
distribution agree very well. The UV continuum is mostly given by numerous iron
lines, and
observations and theory agree particularly well in the
region of the HST/STIS spectrum.

We detected the variability of numerous strong silicon UV lines and proposed
that this is caused by
abundance spots on the stellar surface. However, the overall agreement between
predicted and observed line profiles is poor. Many UV silicon lines are much
weaker than would correspond to the abundances deduced from the optical
spectrum. The variable blueshifted resonance lines of \ion{C}{iv} most likely
originate in the stellar wind, whose mass loss rate (per unit of surface area)
is spatially variable due to inhomogeneous elemental surface distribution. We
found no clear evidence for the magnetically confined 
circumstellar clouds,
whose presence can be detected (from the UV spectra) only when comparing the
observed spectral line shapes with those predicted using detailed
abundance maps of the corresponding element.

\begin{acknowledgements}
We wish to thank D.~Bohlender for providing us his spectra 
and J.~Kub\'at and M.~Oksala for a discussion of this topic.
This work was supported by the grant GA \v{C}R P209/12/0217.
This research was partly based on the IUE data derived from
the INES database using the SPLAT package.
\end{acknowledgements}



\end{document}